# THEORITICAL INVESTIGATIONS ON THE ELASTIC PROPERTIES OF $BiFeO_3$ USING EOS

S.Gaurav[a]* , S.Shankar Subramanian[a], Satyendra P Singh[a] ,B.S Sharma[b]
[a]Department of Physics , AIAS, Amity Unversity ,Noida 201301 India
[b]Department of Physics , Institute of Basic Sciences, Khandari, Agra 282002 India

**Abstract:** Different equations of state (EOS) have been used to study pressure as a function of volume compressions at a given temperature. The critical test of EOS' for solids under low compressions (<10GPa) by evaluating the pressure-volume derivative properties viz., isothermal bulk modulus and its pressure derivative calculated for $BiFeO_3$. The elastic moduli such as bulk modulus, shear modulus, Young's modulus and Poisson's ratio have been calculated as a function of pressure. The values of elastic moduli have been obtained using compressional wave velocity and shear wave velocity. Both compressional and shear velocities and poisson's ratio are in accordance with available experimental data. It is found that B-M EOS gives satisfactory result which is in good agreement with the Stacey EOS.

*Corresponding Author: S.Gaurav
Phone no: 09990871875
E.mail : gauravsharma30@yahoo.com(S.Gaurav)

**Introduction**

The EOS is fundamentally important in studying the properties of materials at different pressure and at high temperature. The knowledge of the P-V-T EOS of relevant standard materials is one of the most basic information needed for pressure calibration. The B-M and Stacey EOS are intended to account for the volumetric properties of solid whose structural configurations vary with pressure and temperature[1-2]. The BM EOS is derived from the Eulerian strain theory [3] where Stacey relates the reciprocal of K' with the ratio P/K [4]. For performing calculation s with the help of an EOS for a material at high pressures, we need the parameters $K_0, K'_0$, all at zero pressure. EOS can be made by studying the variation of K'=dK/dP with pressure or compression(V/Vo). The P-V relation ships reveal that the volume decreases continuously with the increase in pressure. The bulk modulus also increases in pressure but its pressure derivatives K' decreases with the increase in pressure. By presenting a comprehensive analysis of thermodynamics of solids in the limit of extreme compression, Stacey [5-6] has given some basic criteria are satisfied by EOS for its validity and applicability In the limit if infinite pressure (P →∞) at extreme compression (V→0) the pressure derivative of bilk modulus K' approaches a constant value equal to $K'_\infty$ such that;

$$\frac{1}{K'_\infty} = \left(\frac{P}{K}\right)_\infty \qquad (1)$$

Eq.(1) is satisfied at infinite pressure by all such equations of sate for which $K'_O$ is greater than zero.

Room-temperature multiferroic BiFeO3 (BFO) is one of the most intensively studied materials of the moment. Among multiferroics, Bismuth ferrite BiFeO3 (BFO) is commonly considered as a model system[7], and is perhaps the only material that is both magnetic and ferroelectric with a strong electric polarization at ambient conditions. Bismuth ferrite BiFeO3 (BFO) is commonly considered to be a model system for multiferroics[8], especially for ABO3 perovskites where the ferroelectricity is driven by an A cation with $6s^2$ lone pair electrons. The perovskite BFO is one of the very few robust multiferroics with ferroelectric and antiferromagnetic order well above room temperature: In bulk material BFO has an antiferromagnet Néel temperature $T_N$ of 380 °C and a ferroelectric Curie temperature $T_C$ of 830 °C.[9-10] Even though BFO attracts an important attention, the crystal structures of BFO as a function of temperature and pressure are still debated in the literature [7].

The effect of high pressure on perovskites is more complex. However, experiments on LaAlO3 [11] and later on other perovskites [12–15] revealed that some perovskites decrease their tilt angle and undergo phase transitions to higher-symmetry phases on increasing pressure. A pressure-driven diffuse transition, occurring at room temperature in the 40–50 GPa range, that involves a structural change, loss of magnetic order, and metallization has also been observed. The driving force behind such transformations, tentatively attributed to a spin crossover of $Fe^{3+}$, remains to be clarified..

The purpose of the present study is to asses the validity of some important EOSs .A comparison of the result for P-V relationships, bulk modulus and its pressure derivative has been presented with those obtained from Birch Murnaghan third order EOS and Stacey EOS at less than 10 GPa . In this paper, the EOS has been extended to calculate the theoretical values of both compressional and shear velocities of BiFeO3 and comparing these with the available experimental data [15] using isothermal EOS [3]. The other different elastic parameters viz., young's modulus, shear modulus, and poison's ratio are also determined by using pressure – density relationship for bismuth ferrite. In section 2, we present the analysis for the variation of pressure with compression at room temperature. Analysis of elastic moduli is presented in section 3. Discussion and conclusion are presented in section 4.

**2. Variation of pressure with compression at room temperature**

The vales of P, $K_T$ and $K'_T$ has been obtained as a function of compression $V/V_o$ from two EOS for bismuth ferrite with input parameters as $K_o$=97.3GPa and $K_o'$=4.6GPa[20] for pressure range upto 10GPa. The expressions based on different EOS are given below:

B-M EOS:

$$P = \frac{3}{2} K_0 \left( x^{-7} - x^{-5} \right) \left[ 1 + \frac{3}{4} A_1 \left( x^{-2} - 1 \right) \right] \tag{2}$$

$$K_T = \frac{1}{2} K_0 \left( 7x^{-7} - 5x^{-5} \right) + \frac{3}{8} K_0 A_1 \left( 9x^{-9} - 14x^{-7} + 5x^{-5} \right), \tag{3}$$

$$K'_T = \frac{K_0}{8 K_T} \left[ \left( K'_0 - 4 \right) \left( 81x^{-9} - 98x^{-7} + 25x^{-5} \right) + \frac{4}{3} \left( 49x^{-7} - 25x^{-5} \right) \right] \tag{4}$$

Where $x=(V/V_o)^{1/3}$ and $A_1 = (K'_0-4)$

Stacey EOS

$$\ln\left(\frac{V}{V_0}\right) = \frac{K_o'}{K_\infty'^2}\ln\left(1 - K_\infty'\frac{P}{K}\right) + \left(\frac{K_0'}{K_\infty'} - 1\right)\frac{P}{K} \qquad (5)$$

$$\frac{K}{K_0} = \left(1 - K_\infty'\frac{P}{K}\right)^{-\frac{K_0'}{K_\infty'}} \qquad (6)$$

$$\frac{1}{K'} = \frac{1}{K_0'}\left(1 - \frac{K_\infty'}{K_0'}\right)\frac{P}{K} \qquad (7)$$

The results for P, $K_T$ and $K_T'$ as function of V/Vo down to 0.915 are given in table 1. The results obtained from B-M EOS and Stacey EOS are found to present in general fair agreement with each other.

## 3. Elastic Moduli of BiFeO$_3$

There are mainly two types of sound velocities $V_P$ and $V_S$ i.e for compressional or longitudinal and shear or transverse waves, respectively. These are related to bulk modulus (K), shear modulus (G) and density (ρ as givens below [1,21]) as :

$$V_P = \left(\frac{K + \frac{4}{3}G}{\rho}\right)^{1/2} \qquad (8)$$

$$V_S = \left(\frac{G}{\rho}\right)^{1/2} \qquad (9)$$

The isotropic shear modulus (G) within the voigt limit[1] is as follows:

$$G = \frac{1}{5}(2C_s + 3C_{44}) \quad (10)$$

where $C_s = \frac{1}{2}(C_{11} - C_{12}) \quad (11)$

The adiabatic bulk modulus for cubic solids is:

$$K = \left(\frac{C_{11} + 2C_{12}}{3}\right) \quad (12)$$

The Cauchy relation for elastic constant is:

$$(C_{12} - C_{44}) = 2P \quad (13)$$

Using Eqs (10) to (13), we find

$$G = \frac{3}{5}(K - 2P) \quad (14)$$

Using Eq.(14), we can write the first pressure derivative of shear modulus dG/dP as:

$$G' = \frac{3}{5}(K' - 2) \quad (15)$$

According to static and dynamic elasticity, the Young's modulus (Y) is related to compressional and shear velocities as follows:

$$Y = \rho V_s \left(\frac{4V_s^2 - 3V_p^2}{V_s^2 - V_p^2}\right) \quad (16)$$

We can rewrite Eq.(16) in terms of bulk modulus and shear modulus with help of Equations (8) and (9) as follows:

$$Y = \left(\frac{9KG}{9K + 2G}\right) \quad (17)$$

The Poisson's ratio is determined by the following formula [1]:

$$\sigma = \left(\frac{3K + 4P}{12K - 4P}\right) \quad (18)$$

We make use of Eqs (14) – (18) for calculating shear modulus, its pressure derivatives, Young's modulus and Poisson's ratio with help of K and P as a function of density. The results for these elastic moduli and sound velocities from B-M and Stacey at different compression for BiFeO$_3$ at different compression are presented in table 2 and table 3 respectively. ($\rho_o = 8.17$ Kg/m$^3$ [21])

## 4. Discussion and Conclusion

For determining the values of pressure, isothermal bulk modulus and its derivatives, equations of state 1-6 have been used exclusively. We have theoretically determined the variation of different elastic parameters viz., young's modulus, shear modulus, and poison's ratio of bulk bismuth ferrite with compression. In this study we have employed the equations 7-11 to calculate the young's modulus, shear modulus, poison's ratio and have been reported in table 1 and 2. It has been observed that the pressure increases with decrease in compression of the material as depicted in Fig 1. The pressures calculated are found to be in good agreement with experimental data[20]. It is also observed that the elastic moduli increase with increase in pressure and density. Moreover the density determined from the above calculations mentioned in section 3 increases linearly with the calculated pressures [Fig 2]. The nature of young's modulus, shear modulus, poisson's ratio with compression have been predicted and tabulated. The dependence of sound velocities has been determined using the pressure-density relationship [Fig 3]. Furthermore we have predicted the variation of shear and compression wave velocity with different pressures as depicted in Fig 4.

Our theoretical investigations are in accordance with existing literature and few evidences. The calculated Poisson's ratio is in agreement with current studies [16]. Most practical materials typically have poisson's ratio ν values between 0 and 0.5 [17]. Metal oxides usually have ν values around 0.25[18]. The shear sound velocity has agreed with previous literature [19].The rate of increase of compression velocity with pressure is faster than shear velocity.

The present work has predicted various parameters for low pressures (~10 GPa). Nevertheless the experimental study is under investigation. The two EOS employed in our study are found to be in accordance with each other over the whole analysis including all elastic parameters.

**Table 1: Value of P(GPa), $K_T$ (GPa) and $K_T'$(GPa) calculated from Birch-Murnaghan EOS and Stacey EOS ($K'_\infty$=2.76)**

| $V/V_o$ | $P_{BM}$ (GPa) | $P_{Stacey}$ (GPa) | $K_{T\,BM}$ (GPa) | $K_{T\,Stacey}$ (GPa) | $K_T'{}_{BM}$ (GPa) | $K_T'{}_{Stacey}$ (GPa) |
|---|---|---|---|---|---|---|
| 1.000 | 0.000 | 0.000 | 97.30 | 97.30 | 4.599 | 4.602 |
| 0.999 | 0.098 | 0.098 | 97.75 | 97.75 | 4.594 | 4.596 |
| 0.994 | 0.569 | 0.569 | 99.93 | 99.90 | 4.571 | 4.570 |
| 0.989 | 1.062 | 1.062 | 102.19 | 102.14 | 4.548 | 4.544 |
| 0.984 | 1.577 | 1.577 | 104.54 | 104.45 | 4.525 | 4.518 |
| 0.979 | 2.116 | 2.116 | 106.99 | 106.85 | 4.502 | 4.493 |
| 0.974 | 2.679 | 2.679 | 109.55 | 109.34 | 4.479 | 4.468 |
| 0.969 | 3.269 | 3.268 | 112.20 | 111.92 | 4.457 | 4.443 |
| 0.964 | 3.887 | 3.885 | 114.97 | 114.60 | 4.434 | 4.418 |
| 0.958 | 4.535 | 4.531 | 117.87 | 117.39 | 4.411 | 4.394 |
| 0.953 | 5.214 | 5.208 | 120.88 | 120.28 | 4.388 | 4.370 |
| 0.947 | 5.927 | 5.918 | 124.03 | 123.29 | 4.365 | 4.346 |
| 0.942 | 6.675 | 6.663 | 127.32 | 126.43 | 4.343 | 4.322 |
| 0.936 | 7.460 | 7.444 | 130.76 | 129.69 | 4.319 | 4.299 |
| 0.930 | 8.287 | 8.265 | 134.36 | 133.09 | 4.296 | 4.276 |
| 0.924 | 9.156 | 9.127 | 138.12 | 136.64 | 4.275 | 4.253 |
| 0.918 | 10.070 | 10.034 | 142.07 | 140.33 | 4.251 | 4.231 |
| 0.915 | 11.034 | 10.988 | 143.96 | 144.20 | 4.241 | 4.208 |

**Table 2: Value of elastic moduli calculated from Birch-Murnaghan EOS with different compression for $BiFeO_3$**

| $V/V_o$ | $\rho/\rho_o$ | $\rho$ (g/cc) | $P_{BM}$ (GPa) | G (GPa) | Y (GPa) | $\sigma$ | $V_p$ | $V_s$ |
|---|---|---|---|---|---|---|---|---|
| 1.000 | 1.000 | 8.170 | 0.000 | 58.380 | 145.950 | 0.2500 | 4.630 | 2.673 |
| 0.999 | 1.001 | 8.178 | 0.098 | 58.532 | 146.378 | 0.2504 | 4.636 | 2.675 |
| 0.994 | 1.006 | 8.218 | 0.569 | 59.272 | 148.463 | 0.2524 | 4.666 | 2.686 |
| 0.989 | 1.011 | 8.258 | 1.062 | 60.039 | 150.619 | 0.2543 | 4.697 | 2.696 |
| 0.984 | 1.016 | 8.300 | 1.577 | 60.832 | 152.850 | 0.2563 | 4.729 | 2.707 |
| 0.979 | 1.021 | 8.342 | 2.116 | 61.657 | 155.164 | 0.2583 | 4.762 | 2.719 |
| 0.974 | 1.026 | 8.386 | 2.679 | 62.513 | 157.567 | 0.2603 | 4.796 | 2.730 |
| 0.969 | 1.032 | 8.431 | 3.269 | 63.399 | 160.052 | 0.2623 | 4.830 | 2.742 |
| 0.964 | 1.038 | 8.478 | 3.887 | 64.319 | 162.631 | 0.2642 | 4.866 | 2.754 |
| 0.958 | 1.044 | 8.525 | 4.535 | 65.278 | 165.315 | 0.2662 | 4.902 | 2.767 |
| 0.953 | 1.050 | 8.574 | 5.214 | 66.271 | 168.093 | 0.2682 | 4.940 | 2.780 |
| 0.947 | 1.056 | 8.625 | 5.927 | 67.308 | 170.993 | 0.2702 | 4.978 | 2.794 |
| 0.942 | 1.062 | 8.677 | 6.675 | 68.382 | 173.995 | 0.2722 | 5.018 | 2.807 |
| 0.936 | 1.069 | 8.731 | 7.460 | 69.504 | 177.129 | 0.2742 | 5.059 | 2.822 |
| 0.930 | 1.075 | 8.786 | 8.287 | 70.673 | 180.392 | 0.2762 | 5.101 | 2.836 |
| 0.924 | 1.082 | 8.843 | 9.156 | 71.887 | 183.779 | 0.2782 | 5.144 | 2.851 |
| 0.918 | 1.090 | 8.901 | 10.070 | 73.160 | 187.326 | 0.2802 | 5.188 | 2.867 |
| 0.915 | 1.093 | 8.929 | 11.034 | 73.138 | 187.638 | 0.2828 | 5.200 | 2.862 |

**Table 3: Value of elastic moduli calculated from Stacey EOS with different compression for BiFeO$_3$**

| V/V$_o$ | ρ/ρ$_o$ | ρ (g/cc) | P$_{Stacey}$ (GPa) | G (GPa) | Y (GPa) | σ | V$_p$ | V$_s$ |
|---|---|---|---|---|---|---|---|---|
| 1 | 1.000 | 8.170 | 0.000 | 58.380 | 145.950 | 0.2500 | 4.630 | 2.673 |
| 0.999 | 1.001 | 8.178 | 0.098 | 58.532 | 146.379 | 0.2504 | 4.636 | 2.675 |
| 0.99421 | 1.006 | 8.218 | 0.569 | 59.259 | 148.431 | 0.2524 | 4.666 | 2.685 |
| 0.98934 | 1.011 | 8.258 | 1.062 | 60.008 | 150.541 | 0.2543 | 4.696 | 2.696 |
| 0.98439 | 1.016 | 8.300 | 1.577 | 60.778 | 152.714 | 0.2563 | 4.727 | 2.706 |
| 0.97935 | 1.021 | 8.342 | 2.116 | 61.571 | 154.951 | 0.2583 | 4.759 | 2.717 |
| 0.97422 | 1.026 | 8.386 | 2.679 | 62.388 | 157.255 | 0.2603 | 4.791 | 2.728 |
| 0.96901 | 1.032 | 8.431 | 3.268 | 63.230 | 159.630 | 0.2623 | 4.824 | 2.739 |
| 0.96371 | 1.038 | 8.478 | 3.885 | 64.099 | 162.078 | 0.2643 | 4.858 | 2.750 |
| 0.95831 | 1.044 | 8.525 | 4.531 | 64.995 | 164.604 | 0.2663 | 4.892 | 2.761 |
| 0.95283 | 1.050 | 8.574 | 5.208 | 65.919 | 167.212 | 0.2683 | 4.927 | 2.773 |
| 0.94724 | 1.056 | 8.625 | 5.918 | 66.874 | 169.904 | 0.2703 | 4.963 | 2.785 |
| 0.94157 | 1.062 | 8.677 | 6.663 | 67.861 | 172.686 | 0.2724 | 5.000 | 2.797 |
| 0.93579 | 1.069 | 8.731 | 7.444 | 68.881 | 175.562 | 0.2744 | 5.037 | 2.809 |
| 0.92991 | 1.075 | 8.786 | 8.265 | 69.936 | 178.537 | 0.2764 | 5.075 | 2.821 |
| 0.92394 | 1.082 | 8.843 | 9.127 | 71.029 | 181.615 | 0.2785 | 5.115 | 2.834 |
| 0.91785 | 1.090 | 8.901 | 10.034 | 72.160 | 184.804 | 0.2805 | 5.155 | 2.847 |
| 0.915 | 1.093 | 8.929 | 10.988 | 73.332 | 188.108 | 0.2826 | 5.205 | 2.866 |

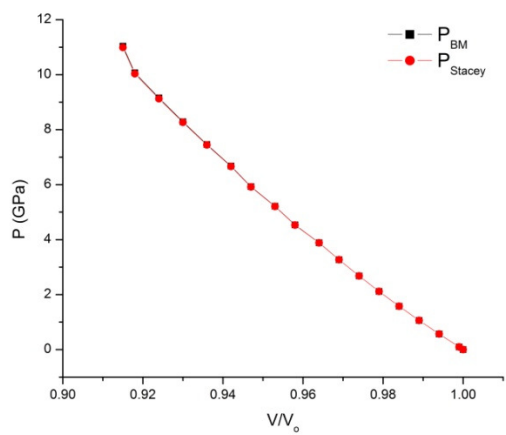

Fig 1.Variation of Pressure with Compression

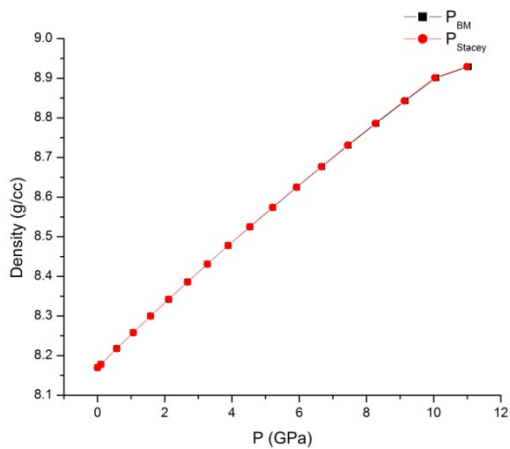

Fig 2.Variation Density with Pressure

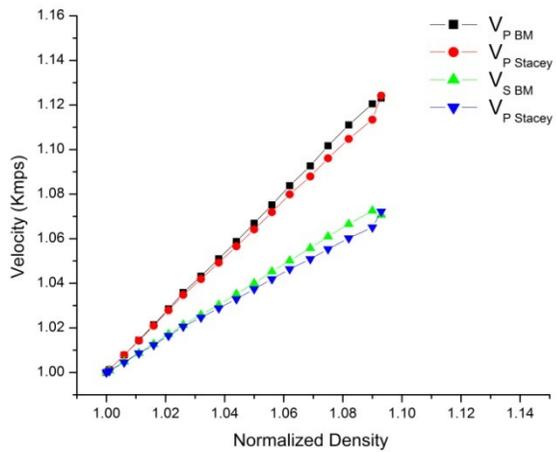

Fig 3.Variation of Reduced velocities with normalized density

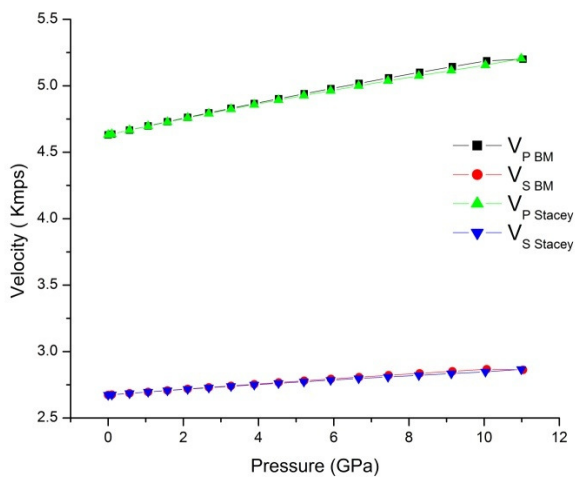

Fig 4. Variation of velocity with Pressure